\documentclass[aps,prl,amssymb,amsmath,%
superscriptaddress,reprint,showpacs]{revtex4-1}

\usepackage[english]{babel}
\usepackage[utf8]{inputenc}

\usepackage[dvips]{graphicx}
\usepackage[breaklinks]{hyperref}
\usepackage[sort&compress]{natbib}

\DeclareMathOperator{\Tr}{Tr}

\begin{document}

\title{Phase Structure of QCD Matter in a Chiral Effective Model with
  Quarks}

\author{Philip Rau}%
\email[]{rau@th.physik.uni-frankfurt.de}%
\affiliation{Institut f\"ur Theoretische Physik, Goethe Universit\"at,
  Max-von-Laue-Str.\ 1, 60438 Frankfurt am Main, Germany}%
\affiliation{Frankfurt Institute for Advanced Studies (FIAS),
  Ruth-Moufang-Str.\ 1, 60438 Frankfurt am Main, Germany}%

\author{Jan Steinheimer}%
\affiliation{Frankfurt Institute for Advanced Studies (FIAS),
  Ruth-Moufang-Str.\ 1, 60438 Frankfurt am Main, Germany}%

\author{Stefan Schramm}%
\affiliation{Institut f\"ur Theoretische Physik, Goethe Universit\"at,
  Max-von-Laue-Str.\ 1, 60438 Frankfurt am Main, Germany}%
\affiliation{Frankfurt Institute for Advanced Studies (FIAS),
  Ruth-Moufang-Str.\ 1, 60438 Frankfurt am Main, Germany}%


\begin{abstract}
  Using a unified hadron-quark effective model for the QCD equation of
  state, this paper studies the phase structure of strongly
  interacting matter in a wide range of temperature and baryonchemical
  potential. At small potentials the model yields a smooth cross-over
  to chirally restored matter with a transition temperature and
  curvature in line with recent lattice QCD estimates and thermal
  model fits of freeze-out curves. Trajectories of constant entropy
  per net baryon number show a clear dependence on the particle
  composition in the model and on repulsive vector field
  interactions. Although the model might feature a critical end-point
  at a rather high baryonchemical potential and low temperature,
  probing it in heavy-ion collisions might be highly challenging due
  to a large thermodynamic spread of matter in the collision fireball.
\end{abstract}

\maketitle 

\section{Introduction}
\label{sec:introduction}
A main objective in studying relativistic heavy-ion collisions lies in
investigating the behavior of strongly-interacting matter under
extreme conditions, i.e.\ high temperatures $T$ and baryon densities
$\rho_B$. Particularly focusing on the phase structure and on mapping
the phase transitions to chirally restored and deconfined matter in
the QCD phase diagram. Experimentally this can accomplished by
studying observables for the phase transition at different beam
energies corresponding to different excitation energies and baryon
densities. While results from high-energy collisions at RHIC suggest
the existence of a quark-gluon plasma (QGP) at high $T$ and small
$\mu_B$~\cite{adams_experimental_2005, adcox_formation_2005}, the QCD
phase structure in other regions (i.e.\ $\mu_B > 0$) remains largely
unknown. Searching for transition signatures by scanning a range of
beam energies was part of the SPS program~\cite{Alt:2007aa}, is
currently performed at RHIC~\cite{Shi:2012ba}, and will also be a key
objective of the CBM experiment at FAIR~\cite{Hohne:2009yw}.\par
At $\mu_B = 0$ lattice QCD consistently shows a smooth cross-over
transition with a ``critical'' temperature $T_c \approx
155$~MeV~\cite{aoki_order_2006, cheng_transition_2006}. At $\mu_B > 0$
lattice QCD standard lattice QCD methods fail due to the fermion sign
problem and estimates from extrapolation are ambiguous. While older
results suggest a critical end-point (CEP) with shift from a smooth
cross-over to a first-order phase transition at finite
$\mu_B$~\cite{Halasz:1998qr, de_forcrand_qcd_2003,
  fodor_critical_2004}, more recent continuum extrapolated lattice
estimates do not necessarily show a first order phase
transition~\cite{Borsanyi:2012cr}. So far there are no experimental
indications for a CEP to exist.\par
Further information on the phase structure of QCD matter is provided
by effective models such as pure quark Polyakov-loop extended
Nambu-Jona-Lasinio (PNJL) models~\cite{fukushima_chiral_2003,
  meisinger_coupling_1995, ratti_phases_2006, ratti_phase_2006,
  roessner_polyakov_2006} or Polyakov-quark-meson (PQM)
models~\cite{Schaefer:2007pw, Herbst:2010rf, Skokov:2010wb,
  Haas:2013qwp}. However, since in these models the Polyakov loop
potentials are fixed at vanishing $\mu_B$~\cite{Schaefer:2007pw},
their validity for describing QCD matter decreases with higher
potentials and baryon densities. To the disadvantage of these models,
in the high-$\mu_B$ region, baryon densities become large and baryon
resonances may exhibit high multiplicities and tend to affect the
phase structure significantly~\cite{rau_baryon_2012}. To circumvent
these restraints, in this work the QCD phase structure is studied
using a unified chiral effective model which combines hadron and quark
degrees of freedom in a single partition function and provides the
correct degrees of freedom in a wide range of $T$ and $\mu_B$.

\section{Model}
\label{sec:model}
This study of the QCD phase structure uses a chiral SU(3)-flavor
$\sigma$-$\omega$ model~\cite{boguta_systematics_1983,
  Papazoglou:1997uw, Papazoglou:1998vr, dexheimer_proto-neutron_2008,
  rau_chiral_2013} for describing the hadronic phase and a PNJL-type
approach for quarks; see~\cite{rau_chiral_2013} for a detailed
review of the model and all parameters. Particles in the model include
all baryons from the octet, the decuplet and all known resonances with
$m \le 2.6$~GeV~\cite{particle_data_group_review_2012}, as well as the
full set of scalar, pseudoscalar, vector, and axial vector mesons
including all meson resonance states. Additionally, the three lightest
quark flavors ($u$, $d$, $s$) are included. The fields are treated in
mean field approximation and correspond to the chiral quark
condensates, i.e.\ the scalar $\sigma$, its strange counterpart
$\zeta$, and the vector $\omega$ and $\phi$ fields. The interaction
between particles and fields is described by
\begin{equation}
  \label{eq:L_BM}
  \mathcal{L}_{\rm int} = -\sum_i \bar{\psi_i} \left[ \gamma_0 \left(
      g_{i\omega} \omega^0 + g_{i\phi} \phi^0 \right) + m_i^* \right]
  \psi_i,
\end{equation}
with $i$ running over all baryons and quarks. Scalar field couplings
$g_{i \sigma, \zeta}$ dynamically generate the effective masses
\begin{equation}
  \label{eq:effective_mass}
  m_i^* = g_{i\sigma}\sigma + g_{i\zeta} \zeta + \delta m_i ,
\end{equation}
except for a small explicit mass $\delta m_i$ ($\delta m_{u, d} =
6$~MeV, $\delta m_s = 105$~MeV for quarks and $\delta m_{n, p} =
150$~MeV for nucleons). The dynamic mass generation ensures decreasing
masses with higher $T$ and $\mu_B$ and, thus, the restoration of
chiral symmetry. The vector couplings $g_{i \omega, \phi}$ generate
the effective chemical potentials accordingly
\begin{equation}
  \label{eq:effective_pot}
  \mu^*_i = \mu_i - g_{i \omega} \omega - g_{i \phi} \phi.
\end{equation}
Using the notation $ X = \sigma^2 + \zeta^2 $, the scalar meson
self-interactions are introduced as
\begin{align}
  \begin{split}
    \label{eq:L_scal}
    \mathcal{L}_{\rm scal} &= -\frac{1}{2} k_0 \, X + k_1 \, X^2 + k_2
    \, \left( \frac{\sigma^4}{2} + \zeta^4 \right) + k_3
    \,\sigma^2 \zeta \\
    &\quad - k_4 \, \chi^4 - \frac{1}{4} \chi^4\, \ln
    \frac{\chi^4}{\chi_0^4} + \frac{\delta}{3}\, \chi^4 \ln{
      \frac{\sigma^2 \zeta} {\sigma_0^2 \zeta_0} },
  \end{split}
\end{align}
with the last two terms describing QCD trace anomaly by introducing
the gluon condensate $\chi$ (dilaton
field)~\cite{Papazoglou:1998vr}. The vector meson
self-interactions are given by
\begin{align}
  \begin{split}
    \label{eq:L_vec}
    \mathcal{L}_{\rm vec} &= \frac{1}{2} \frac{\chi}{\chi_0} \left(
      m^2_{\omega} \omega^2 + m^2_{\phi} \phi^2 \right)\\
    & \quad +g_4 \left( \omega^4 + \frac{\phi^4}{4} + 3 \omega^2
      \phi^2 + \frac{4 \omega^3 \phi}{\sqrt{2}} + \frac{2 \omega
        \phi^3}{\sqrt{2}} \right)
  \end{split}
\end{align}
and the explicit breaking of chiral symmetry due to non-zero current
quark masses adds the following terms to the Lagrangian
\begin{equation}
    \label{eq:self_int_vec_mes}
    \mathcal{L}_{\rm SB} = - \frac{\chi^2}{\chi_0^2} \left[ m_\pi^2
      f_\pi\sigma+\left( \sqrt{2}m_k^ 2f_k -\frac{1}{\sqrt{2}}m_\pi^ 2
        f_\pi \right) \zeta \right] .
\end{equation}
The baryon couplings to the fields are fixed such as to reproduce
nuclear saturation properties and vacuum
masses~\cite{Papazoglou:1997uw, Papazoglou:1998vr} and the quark
couplings are chosen according to the additive quark model and to
avoid free quarks from appearing in the ground state. All baryon
resonance couplings are scaled to the respective nucleon couplings via
$g_{B_i \sigma, \zeta } = r_{s} \, g_{N \sigma, \zeta}$ and $g_{B_i
  \omega, \phi} = r_{v} \, g_{N \omega,
  \phi}$~\cite{rau_baryon_2012}. While the scalar coupling stays $r_s
\approx 1$ to obtain a smooth cross-over at $\mu_B = 0$, the vector
coupling $r_v$ is a free parameter. The repulsive effect of the vector
couplings controls the particle abundances at finite $\mu_B$
(Eq.~\eqref{eq:effective_pot}) and, thus, has large impact on the
resulting phase structure. In the hadron sector, reasonably large
vector couplings ($r_v \approx 1$) cause the disappearance of a
first-order phase transition but yield a smooth cross-over transition
due to the gradual appearance of baryon resonances with higher $T$ and
$\mu_B$~\cite{rau_baryon_2012}. In contrast, all quark vector
couplings have to vanish in order not to fully quench baryon number
fluctuations in the transition
region~\cite{Rau:2013xya}.\par
Quarks are introduced as in PNJL models, defining the scalar Polyakov
loop field $\Phi$ by tracing the time component $A_0$ of the SU(3)
color gauge field $\Phi = 1/3\, \Tr{\left[ \exp{\left( - A_0 / T
      \right) } \right] }$. For static quark masses, $\Phi$ is an
order parameter for deconfinement indicating the breakdown of Z(3)
center symmetry. The effective Polyakov loop potential
\begin{align}
  \label{eq:Polyakov-loop-eff-pot}
  \begin{split}
    U = - \left( a(T) \bar{\Phi} \Phi \right) / 2 + b \left( T_0 / T
    \right)^3 \ln \left[ 1 - 6\bar{\Phi}\Phi \right.\\
    \left. + 4 ( \bar{\Phi}^3 + \Phi^3 ) - 3 ( \bar{\Phi}\Phi )^2
    \right],
  \end{split}
\end{align}
with $ a(T) = a_0 + a_1 \left( T_0/ T \right) + a_2 \left( T_0/T
\right)^2$ and the critical Polyakov temperature $T_0$ is taken
from~\cite{ratti_thermodynamics_2006}. It enters the grand canonical
potential and controls the transition from hadrons to quarks. It is
constructed such as to reproduce quenched lattice QCD thermodynamics
and known features of the deconfinement
transition~\cite{ratti_thermodynamics_2006}. In the confined phase,
the minimum of $U(T,\Phi,\bar{\Phi})$ is located at $\Phi = 0$ and it
moves towards $\Phi \rightarrow 1$ with increasing $T$. Furthermore,
$\Phi$ couples to the dilaton field 
\begin{equation}
  \label{eq:chi-coupling}
  \chi = \chi_0\left[ 1 - 1/4\, \left( \Phi^2 + \bar{\Phi}^2 \right)^2
  \right]
\end{equation}
to suppress the chiral condensate in the quark phase.\par
All thermodynamic quantities are derived from the grand canonical
potential
\begin{equation}
  \label{eq:grand_canon_pot}
  \Omega / V = -\mathcal{L}_{\rm int} - \mathcal{L}_{\rm mes}
  + \Omega_{\rm th} / V  - U_{\rm Pol},
\end{equation}
with $\Omega_{\rm th}$ including thermal contributions from mesons,
baryons, and quarks ($j = u,d,s$) in the form
\begin{align}
  \label{eq:gc_pot_thermal_qq}
  \Omega_{\rm q\bar{q}} = &- T \sum_j \frac{\gamma_j}{(2
    \pi)^3} \int d^3k \; \left( \ln \left[ 1 + \Phi\, e^{ -\frac{1}{T}
        \left( E^*_j(k) - \mu^*_j
        \right)}\right] \right.\nonumber \\
  &\left. +\ln \left[ 1 + \bar{\Phi}\, e^{ -\frac{1}{T} \left(
          E^*_j(k) + \mu^*_j \right)} \right] \right),
\end{align}
with the spin-isospin degeneracy factor $\gamma_j$, and the single
particle energy $E^*_j(k) = \left( k^2 + m_j^{*2} \right)^{1/2}$. By
minimizing $\Omega / V (T,\mu)$ with respect to the fields one obtains
the self-consistent equations of motion for fields and particle
densities. From these, thermodynamic variables are derived via the
pressure $p = -\partial \Omega / \partial V$, the entropy density $s
= \partial p / \partial T $, and the energy density $\varepsilon = Ts
- pV + \sum_i \mu_i \,\rho_i$.\par
A shift in the degrees of freedom from a hadron resonance gas (HRG) at
low $T$ and $\rho_B$ to a pure quark gas in the the high-$T$,
high-$\rho$ region is attained by including an eigenvolume $V_{\rm
  ex}^i$ for hadrons~\cite{rischke_excluded_1991,
  cleymans_excluded_1992, steinheimer_effective_2011}. The baryon
$V_{\rm ex}^B$ is chosen according to the proton charge
radius~\cite{mohr_2010_2011}, mesons are assumed to exhibit half of
this radius, and quarks remain point-like. At high $T$ and $\mu$, when
quark multiplicities rise quickly, this formalism suppresses hadrons
and establishes the shift to a pure quark phase. Thermodynamic
consistency is preserved by redefining $\mu^*_i$, i.e.\ reducing it by
the occupied volume~\cite{rau_chiral_2013}, and multiply the particle
densities $\rho_i$ as well as $\varepsilon$ and $s$ with a correction
factor given by the ratio of the total volume to the non-occupied
sub-volume.\par

\section{Results}
\label{sec:results}
\begin{figure}[tb]
  \centering
  \includegraphics[width=1.\columnwidth]{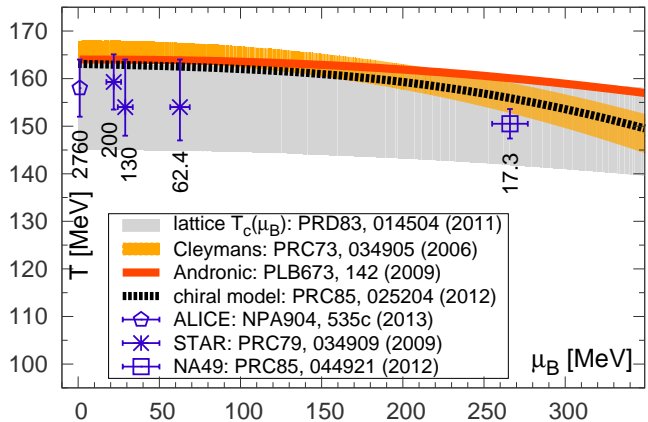}
  \caption{Chiral transition at small $\mu_B$ from lattice
    QCD~\cite{kaczmarek_phase_2011, bazavov_chiral_2011} (gray band)
    and from the chiral model~\cite{rau_baryon_2012} (black line)
    contrasted to freeze-out curves from statistical and thermal model
    fits~\cite{cleymans_comparison_2006, Zschiesche:2006rf,
      andronic_thermal_2009, Abelev:2008ab,
      becattini_hadronization_2012, Andronic:2012dm} for SPS to LHC
    energies ($\sqrt{s_{\rm NN}}$ in GeV).}
  \label{fig:PD}
\end{figure}
The chiral transition extracted from the hadron sector of the
model~\cite{rau_baryon_2012} (black line in Fig.~\ref{fig:PD}) can be
parametrized analog to a lattice QCD
estimate~\cite{kaczmarek_phase_2011} by
\begin{equation}
  \label{eq:chiral-transition-curve-HQ-model}
  T_c(\mu_B) = T_0 \left( 1 -
    0.0193\left( \frac{\mu_B}{T_0} \right)^2 \right)
\end{equation}
with $T_0 = T_c(\mu_B = 0) = 164$~MeV. In the chiral model this curve
is extracted at the point of the steepest decrease in $\sigma(T)$ at a
given baryonchemical potential. The curve agrees well with different
models and experimental results at different beam
energies. In~\cite{Tawfik:2013eua} a constant value of the interaction
measure $(e-3p) / T^4 = 7/2$ was proposed to reliably parametrize the
chemical freeze-out curve. This parametrization can be reproduced in
the hadronic sector of the chiral model. However, since in the chiral
transition region the interaction measure increases very fast with
higher $T$, other constant values of $(e-3p) / T^4$ close to $7/2$
also yield curves close to recent freeze-out parametrizations.  In
general, the transition behavior is not affected by additionally
taking into account the quark phase. Even in the presence of a quark
phase, baryon multiplicities at the chiral transition are high and the
steepest decline in $\sigma$ still takes place in the HRG. But when
considering quarks, $T_c$ is shifted to slightly higher
values~\cite{rau_chiral_2013}. For this reason, and due to the
non-existence of direct coupling effects between $\Phi$ and the quark
condensates, quarks have only minor impact on the critical temperature
of the chiral transition.\par
In~\cite{Randrup:2006uz} it is shown that, using a fixed target setup,
highest net baryon number densities at freeze-out can be achieved in
heavy-ion collisions at energies close to $E_{\rm lab} =
30$A~GeV. Accelerators at the upcoming FAIR facility, operating
between $5$ and $40$~GeV per nucleon, will cover this energy range.
\begin{figure}[tb]
  \centering
  \includegraphics[width=1.\columnwidth]{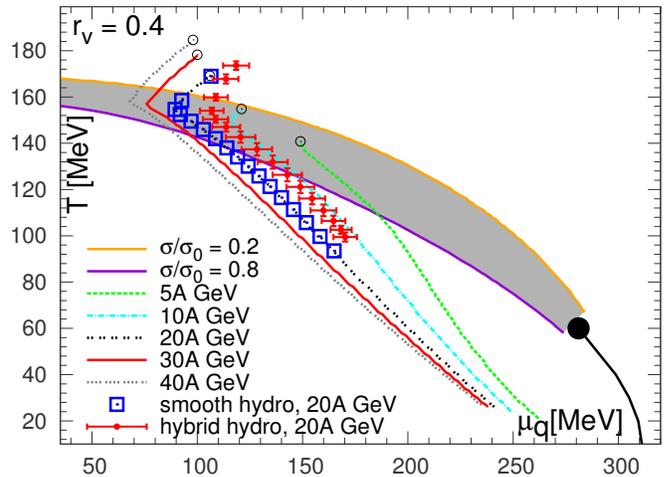}
  \caption{Isentropic expansion paths (lines of constant $S/A$) for
    different $E_{\rm lab}$ in the $T$--$\mu_q$-plane of the pure
    hadron EoS ($r_v = 0.4$) together with lines of constant
    $\sigma/\sigma_0$ and the first-order phase transition with a CEP
    (black solid line). Red points depict the central cell in the
    UrQMD hybrid model (Au+Au at $E_{\rm lab} = 20$A~GeV, averaged
    over 400 events) and blue squares illustrate a smooth hydrodynamic
    evolution with a time interval of 1~fm between the data points.}
  \label{fig:isen_rv04}
\end{figure}
Figure~\ref{fig:isen_rv04} and~\ref{fig:isen_rv09} show the isentropic
expansion paths, i.e.\ lines of constant entropy per net baryon number
$S/A$, corresponding to these collision energies. The adiabats are
depicted in the hadronic equation of state (EoS) without a quark phase
and feature two values for the resonance vector couplings. In case of
a rather weak coupling $r_v = 0.4$ (Fig.~\ref{fig:isen_rv04}) the
phase structure exhibits a first order phase transition up to $T
\approx 60$~MeV, an adjacent CEP, and a smooth but rapid cross-over
transition for higher temperatures (gray band). In the transition
region, with increasing $T$, the rapid incline in $m_i^*$ of the
baryons induces large baryon multiplicities and a sharply rising
energy density and pressure along the $\mu_q$-axis (i.e.\ $\mu_q =
\mu_B/3$) causing the adiabats to bend sharply at the chiral
transition distant to the CEP.\par
\begin{figure}[tb]
  \centering
  \includegraphics[width=1.\columnwidth]{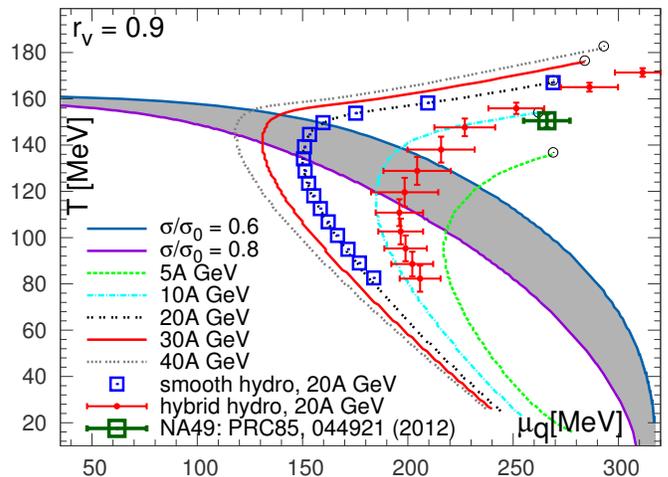}
  \caption{Adiabats as in Fig.\ref{fig:isen_rv04} using a larger
    resonance vector coupling $r_v = 0.9$, for which no first order
    phase transition exists. The green marker depicts the freeze-out
    point extracted from NA49 data~\cite{becattini_hadronization_2012}
    (cf.\ Fig.~\ref{fig:PD}).}
  \label{fig:isen_rv09}
\end{figure}
This behavior changes for more reasonable resonance vector couplings
$r_v = 0.9$ (Fig.~\ref{fig:isen_rv09}) for which the changes in
$\varepsilon$ and $p$ are much slower along the $\mu_q$-axis and the
chiral transition takes place in a much broader $T$-range due
suppressive vector field interactions. In this case, neither a first
order phase transition nor a CEP exists and the chiral transition
takes place in a much broader $T$-range. Compared to $r_v = 0.4$,
substantial softening of the EoS with larger $r_v$ causes the adiabats
to smoothly bend at the chiral transition and to reach notably higher
$\mu_q$ while the initial $T$ only changes on a minor scale.\par
The figures also depict dynamic expansion paths of the fireball at
20A~GeV from ideal hydrodynamics~\cite{Rischke:1995ir} using initial
densities from a geometric overlap model~\cite{Reiter:1998uq} (blue
squares). Also shown are collision dynamics from the UrQMD hybrid
model with fluctuating initial conditions~\cite{Petersen:2008dd} (red
crosses) sampling over the central cell in 400 Au+Au collisions with a
time interval of $\Delta t = 1$~fm between data points. In case of a
soft EoS (Fig.~\ref{fig:isen_rv09}), initial fluctuations cause a
larger dispersion in $T$ and $\mu$ on an event-by-event
basis. Furthermore, initial state density fluctuations create
sub-regions in the fireball and cause a notable dispersion of the
thermodynamic properties of the fireball matter. Due to this effect,
QCD matter may spread over at least 50~MeV in $\mu_q$ in each single
event~\cite{Bass:2012gy}. This means that there is no well defined and
narrow isentropic adiabat corresponding to a single event and energy
but rather a large area of the phase diagram is covered in one
collision at a specific energy. While on the one hand, this eventually
allows for probing regions well outside a narrow adiabat, on the other
hand poses the question of how to map back final state observables to
matter properties to one specific point in $T$ and $\mu$ on the QCD
phase diagram.\par
\begin{figure}[tb]
  \centering
  \includegraphics[width=1.\columnwidth]{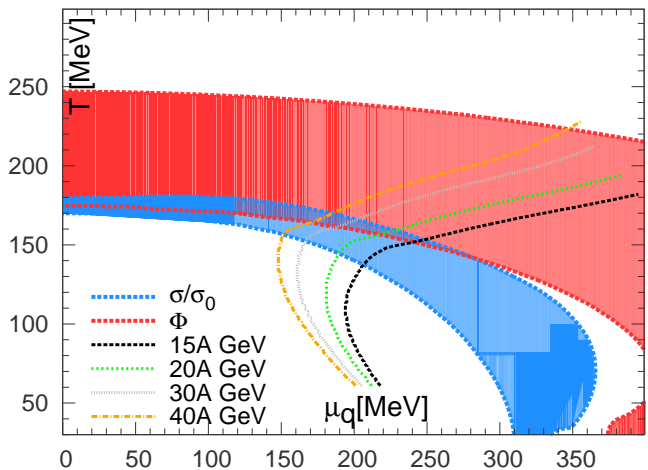}
  \caption{Order parameter $\sigma/\sigma_0$ (blue) and $\Phi$ (red)
    of the effective model including hadrons and quarks along with
    isentropic adiabats. In the colored areas $\sigma/\sigma_0$
    declines from $0.7$ to $0.4$ and $\Phi$ rises from $0.7$ to $0.4$
    with increasing $T$ and $\mu_q$. The vector couplings are
    $g_{q\omega} = 0$ for quarks and $r_v = 0.9$ for baryon
    resonances.}
  \label{fig:isen_hq}
\end{figure}
Additionally including quarks (Fig.~\ref{fig:isen_hq} using
$g_{q\omega} = 0$) has only minor effect on the adiabat curvature at
the chiral transition. However, in the presence of a quark phase, the
adiabats are slightly steeper in the chiral limit and higher initial
$T$ and $\mu_q$ are achieved. In close analogy to the effect of the
vector coupling of resonances $r_v$ in the purely hadronic EoS,
turning up the quark vector coupling $g_{q\omega}$ from zero to finite
values, causes significantly higher initial values in $\mu_q$ and
almost flat adiabat curvatures above the chiral transition in the full
model including quarks.

\section{Summary}
\label{sec:summary}
In summary, using an effective model with hadrons and quarks, we
present a parametrization for the chiral transition at small $\mu_B$
agreeing well with other recent results on this topic. Isentropic
adiabats in the EoS are discussed in the context of varying vector
interactions: Probing the CEP and a first order phase transition at
FAIR energies might be difficult due to the distance of isentropic
adiabats to the first order phase transition in the
model. Additionally, probing specific points in the QCD phase diagram
may be even more complicated considering the sizable thermodynamic
spread of fireball matter induced by initial state fluctuations as
seen in dynamic models for heavy-ion collisions.\par

\section{Acknowledgements}
\label{sec:ack}
This work was supported by BMBF, GSI, and the Hessian LOEWE initiative
through the Helmholtz International Center for FAIR and the Helmholtz
Graduate School for Hadron and Ion Research. Computing resources were
provided by the Center for Scientific Computing of the Goethe
University Frankfurt.

\bibliographystyle{apsrev}
\bibliography{Bib}

\end{document}